\documentclass[aps,prc,reprint,superscriptaddress,nofootinbib]{revtex4-1}
\usepackage{graphicx}
\usepackage{multirow}
\usepackage{comment}
\usepackage{color}
\usepackage{amssymb}
\usepackage{amsmath}
\usepackage{lineno}
\def\nuc#1#2{\relax\ifmmode{}^{#1}{\protect\text{#2}}\else${}^{#1}$#2\fi}

\newcommand{\vecr}{{\vec r}}

\newcommand{\be}{\begin{eqnarray}}
\newcommand{\ee}{\end{eqnarray}}

\newcommand{\bwt}{\begin{widetext}}
\newcommand{\ewt}{\end{widetext}}

\begin{document}
%\begin{CJK*}{GB}{song}
% Use the \preprint command to place your local institutional report
% number in the upper righthand corner of the title page in preprint mode.
% Multiple \preprint commands are allowed.
% Use the 'preprintnumbers' class option to override journal defaults
% to display numbers if necessary
%\preprint{}

%Title of paper

%\title{Interpretation of large $\alpha$-yields in $^{6}$Li induced reactions}  
\title{Comprehensive analysis of large $\alpha$ yields observed in $^{6}$Li induced reactions}

% repeat the \author .. \affiliation  etc. as needed
% \email, \thanks, \homepage, \altaffiliation all apply to the current
% author. Explanatory text should go in the []'s, actual e-mail
% address or url should go in the {}'s for \email and \homepage.
% Please use the appropriate macro foreach each type of information

% \affiliation command applies to all authors since the last
% \affiliation command. The \affiliation command should follow the
% other information
% \affiliation can be followed by \email, \homepage, \thanks as well.
\author{Jin Lei}
\email[]{jinlei@us.es}
\altaffiliation{Present address: Institute of Nuclear and Particle Physics, and Department of Physics and Astronomy, Ohio University, Athens, Ohio 45701, USA}
%\homepage[]{Your web page}
%\thanks{}

\affiliation{Departamento de FAMN, Universidad de Sevilla, 
Apartado 1065, 41080 Sevilla, Spain.}

\author{Antonio M. Moro}
\email[]{moro@us.es}
%\homepage[]{Your web page}
%\thanks{}

\affiliation{Departamento de FAMN, Universidad de Sevilla, 
Apartado 1065, 41080 Sevilla, Spain.}

%Collaboration name if desired (requires use of superscriptaddress
%option in \documentclass). \noaffiliation is required (may also be
%used with the \author command).
%\collaboration can be followed by \email, \homepage, \thanks as well.
%\collaboration{}
%\noaffiliation

\begin{abstract}
\begin{description}
\item[Background] Large $\alpha$ yields have been reported over the years in reactions with $^{6}$Li and $^{7}$Li projectiles. Previous theoretical analyses have shown that the elastic breakup (EBU) mechanism (i.e., projectile breakup leaving the target in its ground state) is able to account only for a small fraction of the total $\alpha$  inclusive breakup cross sections, pointing toward the dominance of  non-elastic breakup (NEB) mechanisms.  
%these $\alpha$ particles stem from several competing mechanisms, such as direct breakup, transfer, and fusion (complete and incomplete).  

\item[Purpose] We aim to provide a systematic study of the $\alpha$ inclusive cross sections observed in nuclear reactions induced by $^{6}$Li projectiles. In addition to estimating the total $\alpha$ singles cross sections, it is our goal to evaluate angular and energy distributions of these $\alpha$ particles and compare with experimental data, when available. 

\item[Method] We compute separately the EBU  and NEB components of the inclusive breakup cross sections. For the former, we use the continuum-discretized coupled-channels (CDCC) method, which treats this mechanism to all orders. For the NEB part, we employ the the model proposed in the eighties by Ichimura, Austern and Vincent [Phys. Rev. C32, 432 (1982)], within the DWBA approximation. 

\item[Results] Overall, the sum of the computed EBU and NEB cross sections is found to reproduce very well the measured singles cross sections. In all cases analyzed, we find that the inclusive breakup cross section is largely dominated by the NEB component. 

\item[Conclusions] The presented method provides a global and systematic description of inclusive breakup reactions induced by $^{6}$Li projectiles. It provides also a natural explanation of the previously observed underestimation of the measured $\alpha$  yields by CDCC calculations. The method used here can be extended to other weakly-bound projectiles, including halo nuclei. 
\end{description} 
\end{abstract}

%\begin{keyword}
%closed form sum rules for inclusive breakup, DWBA, incomplete fusion.
%\end{keyword}  
\pacs{ 25.60.Dz,25.60.Gc,25.60.Bx,21.10.Gv,27.20.+n. }
\date{\today}

\maketitle

%%
%% Start line numbering here if you want
%%
% \linenumbers

%---------------------------------------
\section{Introduction \label{sec:intro}}
%---------------------------------------
Reactions induced by the $^6$Li nucleus have been extensively studied 
giving rise to a large body of experimental data at present. Given its  
marked $\alpha +d$ structure, with a separation energy 
of $1.474$ MeV (to be compared with the single nucleon 
separation energy of 5.39 MeV), one may anticipate that the breakup of this nucleus into 
$\alpha$ and $d$ is a major reaction channel.  In fact, experimental data 
show remarkably large yields of $\alpha$ particles but, contrary to what 
naively expected,  these yields are typically much larger than the 
corresponding $d$ yields.  This suggests that the breakup of the 
$^{6}$Li is not a simple direct breakup mechanism. 

From the theoretical point of view, a proper interpretation of these 
$\alpha$ yields is still lacking. Continuum-discretized coupled-channels 
(CDCC) calculations, which treat the $^{6}$Li breakup as an inelastic excitation to 
the continuum, reproduce successfully the coincidence $\alpha+d$ measurements \cite{Signorini03}
but they largely underestimate the inclusive $\alpha$ cross sections.  
It is worthwhile recalling that the CDCC method provides only the so-called 
{\it elastic breakup} (EBU) component of the total breakup cross section. 
For the reaction of a $^{6}$Li projectile impinging on a target $A$, this 
corresponds to the processes of the form ${\rm ^{6}Li}+A \rightarrow \alpha+d +A_\mathrm{g.s.}$ in 
which the two-projectile clusters survive after the collision and the 
target remains in the ground state.%
\footnote{If a three-body description of the $^{6}$Li is used, $\alpha$+p+n, the three-body breakup mode 
${\rm ^{6}Li}+A \rightarrow \alpha+p+n +A_\mathrm{g.s.}$ would be also part of the elastic breakup channel. Since we resort here 
to a two-body model of $^{6}$Li we include this channel in the NEB part.}%
 Thus, the underestimation of the 
inclusive $\alpha$ yields by the CDCC calculations means that there other 
mechanisms contributing to the inclusive breakup cross section other than the EBU. These 
include the exchange of nucleons between $d$ and $A$, the projectile 
dissociation accompanied by target excitation, and the fusion of $d$ by 
$A$, among others, that we will globally denote as {\it non-elastic breakup} (NEB) channels.  
An explicit account of these process is very challenging due to the huge 
number of accessible final states and the variety of competing different 
mechanisms.

When one is only interested in the evaluation of the singles cross section 
(for example, the energy or angular distribution of $\alpha$ particles),  
rather than on the separate contributing mechanisms,  one may resort to 
the inclusive breakup models proposed in the 1980s and recently reexamined 
by several groups \cite{Jin15,Jin15b,Jin16,Car15,Pot15}. In these models, the 
sum over all the possible final states through which the unobserved fragment 
$d$ may interact with the target is done in a formal way, making use of 
the Feshbach projection formalism \cite{Feshbach:1962vp} and closure. 

In this work, we will show that inclusive $\alpha$ singles cross sections from 
$^{6}$Li-induced reactions can be remarkably well reproduced using the 
inclusive breakup model  proposed by Ichimura, Austern and Vincent (IAV) \cite{Ich85}. 
To our knowledge, this is the first study of this kind providing a 
systematic explanation of these data. 

Although the IAV model provides a common formalism for the calculation 
of the elastic and non-elastic breakup components of the inclusive breakup 
cross section, in our analysis we will employ this model only for the NEB 
part, whereas for the EBU part we will use the continuum-discretized coupled-channels (CDCC) method, which treats breakup to all orders.

The paper is organized as follows. In Sec.~\ref{sec:formalism} 
we give a short overview of the IAV theory, highlighting only its main formulas. 
In Sec.~\ref{sec:ex0} the extension of the formalism to  negative 
deuteron energies (bound states) is discussed. 
%In Sec.~\ref{sec:struct} we provide details on the structure models assumed  for the $^{6}$Li projectiles.  
 In Sec.~\ref{sec:calc}, the formalism is 
applied to describe the $\alpha$ cross sections in several $^6$Li-induced reactions comparing with the available data. 
In Sec.~\ref{sec:nebvstr} the role of the transfer channels on the NEB cross section is discussed.
In Sec.~\ref{sec:6lisys} we investigate the systematic behaviour of the 
inclusive cross section with respect to the incident energy and for all analyzed targets. 
Finally, in Sec.~\ref{sec:sum} we summarize the main results of this work.

%--------------------------------------------------------------------------
\section{The Ichimura, Austern, Vincent (IAV) model \label{sec:formalism}}
%----------------------------------------------------------------------------
In this section we briefly summarize the  model of Ichimura, 
Austern and Vincent (IAV), whose original derivation can be found 
in \cite{Ich85,Aus87}, and has been also recently revisited by several 
authors \cite{Jin15,Jin15b,Car15,Pot15}. We outline here the main results 
of this model, and refer the reader to these references for further 
details on their derivations.

We write the process under study in the form,
\begin{equation}
%a (=b+x) + A \rightarrow b + \textrm{anything} .
a (=b+x) + A \rightarrow b + B^* .
\end{equation}
where the projectile $a$, composed of $b$ and $x$,  collides with a target $A$, emitting $b$ fragments and any other fragments. Thus, $B^*$ denotes any final state of the $x+A$ system.      
%In the spectator model, one assumes that this process is triggerred by the interaction of the unobserved particle ($x$) with the target ($A$). The fragment $b$ interacts also with the target but, non-elastic processes arising from this interaction (e.g.~target excitation), are included only effectively through some optical potential $U_{b}$.
 
\begin{figure}[tb]
\begin{center}
% {\def\svgwidth{0.9\columnwidth}{\input{zrcoor.pdf_tex}} \par}
 {\centering \resizebox*{0.85\columnwidth}{!}{\includegraphics{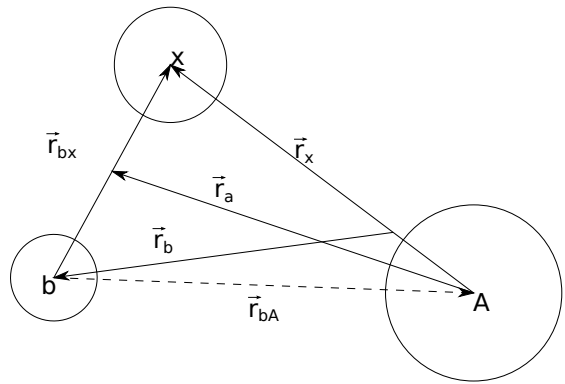}} \par}
\caption{\label{zrcoor}(Color online) Coordinates used in the non-elastic breakup calculations.}
\end{center}
\end{figure}

This process will be described with the effective Hamiltonian
\begin{equation}
\label{eq:H3b}
H= K + V_{bx} + U_{b}(\vecr_{b}) + H_A(\xi) + V_{xA}(\xi,\vecr_{x}) ,
\end{equation}
where $K$ is the total kinetic energy operator, $V_{bx}$ is the 
interaction binding the two clusters $b$ and $x$ in the initial 
composite nucleus $a$, $H_{A}(\xi)$ is the Hamiltonian of the target 
nucleus (with $\xi$ denoting its internal coordinates) and $V_{xA}$ and 
$U_{b}$ are the fragment--target interactions. The relevant coordinates are depicted in Fig.~\ref{zrcoor}. 

In writing the Hamiltonian of the system in the form (\ref{eq:H3b}) we 
make a clear distinction between the two cluster constituents; the 
interaction of the fragment $b$, the one that is assumed to be observed
in the experiment, is described with a (complex) optical potential. Non-elastic 
processes arising from this interaction (e.g.~target excitation), 
are included only effectively through $U_{b}$. The particle $b$ is said 
to act as  {\it spectator}. On the other hand, the interaction of 
the particle $x$ with the target retains the dependence of the target degrees of freedom 
($\xi$).

Starting from Hamiltonian (\ref{eq:H3b}) IAV derived the following expression for the double differential cross section for the NEB  with respect to the 
angle and energy of the $b$ fragments:
\begin{equation}
\label{eq:iav_3b}
\left . \frac{d^2\sigma}{dE_b d\Omega_b} \right |_\mathrm{NEB} = 
-\frac{2}{\hbar v_{a}} \rho_b(E_b)  \langle \psi^{(0)}_x(\vec{k}_b,\vecr_x) 
| W_x | \psi^{(0)}_x (\vec{k}_b,\vecr_x) \rangle   ,
\end{equation}
where $v_a$ is the projectile-target relative velocity,  
$\rho_b(E_b)=k_b \mu_{b} /((2\pi)^3\hbar^2)$ is the density of states 
for the particle $b$,   $W_{x}$ is the imaginary part of the optical 
potential describing $x+A$ elastic scattering and  
$\psi_x^{(0)}(\vec{k}_b,\vecr_x)$ is the so-called $x$-channel 
wave function, which governs the evolution of $x$ after the projectile 
dissociation, when $b$ scatters with momentum $\vec{k}_b$ and the 
target remains in the ground state. This function  satisfies the following 
inhomogeneous differential equation
\begin{equation}
\label{eq:pz_3b}
(E_x - K_x - {U}_{xA})  \psi_x^{(0)}(\vec{k}_b, \vecr_x) =  (\chi_b^{(-)}(\vec{k}_b,\vecr_b) | V_\mathrm{post}|\Psi^{3b} \rangle ,
\end{equation}
where $E_x=E-E_b$,  $\chi_b^{(-)}$ is the distorted-wave describing the 
scattering of $b$ in the final channel with respect to the $x$+$A$ 
sub-system, and $V_\mathrm{post} \equiv V_{bx}+U_{bA}-U_{b}$ 
(with $U_b$ the optical potential in the final channel) is the post-form 
transition operator. The notation $( | |  \rangle$ indicates integration over the $\vecr_b$ coordinte only. This equation is to be solved with outgoing boundary conditions.

Austern {\it et al.} \cite{Aus87} suggest approximating the three-body wave 
function appearing in the source term of Eq.~(\ref{eq:pz_3b}), $\Psi^{3b}$, 
by the CDCC one. Since the CDCC wave function is also a complicated object 
by itself, a simpler choice is to use the DWBA approximation, i.e.,  
$\psi^{3b}_x \approx \chi^{(+)}_{a}(\vecr_a) \phi_{a}(\vecr_{bx})$, 
where $\chi^{(+)}_{a}$ is a distorted wave describing $a+A$ elastic 
scattering and $\phi_{a}$ is the projectile ground state wave function.

The IAV model  has been recently revisited by several groups \cite{Jin15,Pot15,Car15}. 
All the calculations performed so far by these groups make use of the 
DWBA approximation for the incoming wave function.  
In Refs.~\cite{Pot15,Car15}, the theory was applied to deuteron induced 
reactions of the form $A(d,pX)$, and in Ref.~\cite{Jin15} the model was 
extended to $^{6}$Li projectiles, presenting a first application to 
the $^{209}$Bi($^6$Li,$\alpha X$) reaction. 
In general, the agreement with the data has been found to be 
very encouraging, although further comparisons with experimental data 
are advisable to better assess the validity and limitations of the model.

%------------------------------------------------------------
\section{Extension of IAV model to $E_x < 0$}\label{sec:ex0}
%------------------------------------------------------------
The sort of breakup cross section considered by Ichimura, Austern and Vincent can be
regarded as transfer to continuum process populating $x+A$ states with positive relative energy ($E_x > 0$). In general, the inclusive cross section will contain also
 contributions coming from the population of states below the breakup $x+A$ threshold ($E_x < 0$). For example, in a ($^6$Li, $\alpha X$) reaction, the $\alpha$'s
emitted at the higher energies will actually correspond to deuteron transfer to bound states of the
target nucleus. One would like to have a common framework to describe transfer to continuum states
as well as to bound states. The explicit inclusion of all possible final bound states is
unpractical because of their large number and the uncertainties in their spin/parity
assignments and spectroscopic factors. An alternative procedure was proposed by Udagawa and co-workers \cite{Uda89}. 
The key idea is to extend the complex potential to negative energies.  Then, the bound states of the system are simulated by the eigenstates in this complex
potential. The imaginary part will be associated with the spreading width of the single-particle
states, which accounts for the fragmentation of these states into more complicated configurations
due to the residual interactions. The method has been recently reexamined by 
Potel \textit{et al.} \cite{Potel:2015eqa}, who have provided an efficient implementation of this idea. Here, we closely follow their formulation. For that, we first rewrite Eq.~(\ref{eq:pz_3b}) in integral form 
\begin{equation}
\label{eq:phi0_int}
\psi_x^{(0)}(\vec{k}_b, \vecr_x) = \int_0^\infty G_x (\vec{r}_x, \vec{r'}_x) \rho (\vec{k}_b,\vec{r'}_x) \mathrm{d}^3 r_x ,
\end{equation}
where $\rho (\vec{k}_b,\vec{r'}_x)=(\chi_b^{(-)}(\vec{k}_b, \vecr_b) | V_\mathrm{post}|\Psi^{3b} \rangle$ 
is the source term of the inhomogeneous  Eq.~(\ref{eq:pz_3b}) and 
$G_x (\vec{r}_x, \vec{r'}_x)$ is the Green's function  
\begin{equation}
G_x (\vec{r}_x, \vec{r'}_x) = \frac{1}{r_x r'_x} \sum_{l_x m_x} g_{l_x} (r_x,r'_x) Y_{l_x}^{m_x*} (\hat{r'}_x) Y_{l_x}^{m_x} (\hat{r}_x) ,
\end{equation}
where $g_{l_x} (r_x,r'_x)$ satisfies the equation 
\begin{equation}
\label{eq:gfunction}
(E_x - K_x - {U}_{xA}) g_{l_x} (r_x,r'_x) = \delta (r_x-r'_x). 
\end{equation}
As usual, the solution of this equation is obtained from the regular ($f_{l_x}(r_x)$) and irregular ($h_{l_x}^{(+)}(r_x)$)
solutions of the corresponding homogeneous equation. From these two solutions, $g_{l_x} (r_x,r'_x)$ can be expressed as 
\begin{equation}
g_{l_x} (r_x,r'_x) =  N_{l_x} f_{l_x}(r_<) h_{l_x}^{(+)}(r_>) ,
\end{equation}
where $r_<$ is the lesser value of $r_x$ and $r'_x$ and $r_>$ is the larger one. 
The normalization constant $N_{l_x}$ can be found by 
integrating Eq.~(\ref{eq:gfunction}) over an infinitesimal interval around 
$r'_x$
\begin{equation}
\begin{split}
\frac{2 \mu_x}{ \hbar^2}& = \int_{r'_x-\delta}^{r'_x+\delta} 
\mathrm{d}r_x \frac{\mathrm{d}^2}{\mathrm{d}r^2_x} g_{l_x} (r_x,r'_x)= 
\frac{\mathrm{d}}{\mathrm{d}r_x} g_{l_x} (r_x,r'_x)\Bigg|_{r'_x-\delta}^{r'_x+\delta}   \\
& = N_{l_x} \Big[f_{l_x}(r'_x)\frac{\mathrm{d}}{\mathrm{d}r_x}h^{(+)}_{l_x}(r'_x+\delta) \\
 & -  h^{(+)}_{l_x}(r'_x)\frac{\mathrm{d}}{\mathrm{d}r_x}f_{l_x}(r'_x-\delta) \Big] \\
& \xrightarrow{\delta \to 0} N_{l_x}  \mathcal{W}[f_{l_x}(r'_x),h^{(+)}_{l_x}(r'_x)] 
\end{split}
\end{equation}
Where $\mathcal{W}$ denotes a Wronskian, which is independent of the value of $r'_x$. 

It is worth noting that the integral form of the $x-$channel wave function (\ref{eq:phi0_int}) can be also be used for positive $x-A$ energies. Proceeding in this way, the application of the IAV formalism to positive and negative energies is formally analogous. Despite this formal similitude, the interpretation of the channel function and of the underlying imaginary part  of the potential is somewhat different in both regions. For $E_x>0$ the channel function $\psi_x^{(0)}$ describes $x-A$ elastic scattering and the imaginary part is therefore associated with the flux leaving this channel in favor of non-elastic channels. For $E_x<0$, the channel wave-function describes the motion of the $x$ particle in a bound  single-particle configuration state of the residual nucleus, and the imaginary part is connected with the spreading width of this configuration,  which accounts for the fragmentation of these 
states into more complicated configurations. The connection between both regimes becomes more transparent within a dispersive formulation of the optical potential, as suggested long ago by Mahoux and Sartor \cite{Mah86,Mah91} and recently reexamined by several groups (see e.g.~\cite{Dick16}).

\section{Comparison with experimental data \label{sec:calc}}
%------------------------------------------
In this section, we compare the formalism with existing $^{6}$Li inclusive breakup data on different targets. The $^{6}$Li nucleus is treated in a two-cluster model ($\alpha$+$d$), with $\alpha$ and $d$ playing the roles of spectator and participant in the IAV model, respectively.

The elastic breakup (EBU) contribution of the inclusive breakup cross section is evaluated with the CDCC method \cite{Aus87}, using the coupled-channels code {\tt FRESCO} \cite{Thom88}. 
In this method, the breakup is treated as an inelastic excitation to the continuum states of the projectile. 
Although four-body CDCC calculations for $^6$Li scattering have become 
recently available \cite{Wat12}, we rely here on the more conventional 
$\alpha+d$ di-cluster model. Thus, diagonal and off-diagonal coupling potentials 
are generated from the  $d$+target and $\alpha$+target interactions, evaluated 
at 2/3 and 1/3 of the projectile incident energy, respectively.  
In order to reproduce correctly the elastic scattering data, CDCC 
calculations based on this two-body model typically require some renormalization 
of the fragment-target potentials \cite{Hir91,Wat12}. This has been recently found to be a consequence of the shortcomings of the two-body description of the $^{6}$Li nucleus, which results in an effective suppression of the deuteron-target absorption \cite{Wat12}. In our previous work  \cite{Jin15}, we found that this effect could be well simulated  by removing the surface part of the deuteron-target optical potential. In the calculations presented in this work, we also allow for such kind of modification, in order to reproduce 
correctly the elastic scattering data.

For the $\alpha+d$ potential, we use the potential model from 
Ref.~\cite{Nishioka84}, which contains both central and spin-orbit terms, with 
the latter required to place correctly the $\ell=2$ resonances. 

For the non-elastic breakup calculations, we rely also on a  $\alpha+d$ model, but the spin of the deuteron is ignored, since our current implementation of the IAV model ignores the intrinsic spin of the fragments.  This approximation was also used  in our previous works \cite{Jin15,Jin15b,Jin16}.  
%These calculations were performed with the code developed by the authors \cite{Jin16}.

\subsection{$^{208}$Pb ($^6$Li, $\alpha X$) \label{sec:6li208pb}}

%-------------------------------------------------------------
% 6Li+208Pb breakup ds/dw
\begin{figure}[tb]
\begin{center}
 {\centering \resizebox*{0.95\columnwidth}{!}{\includegraphics{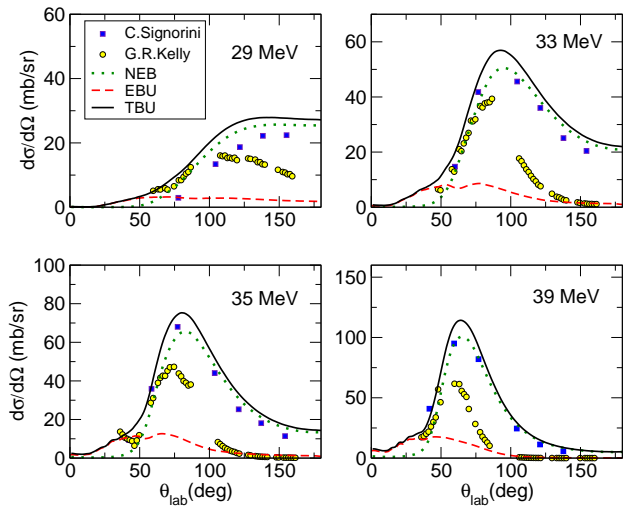}} \par}
\caption{\label{fig:li6pb_dsdw}(Color online) Angular distribution of 
$\alpha$ particles produced in the reaction $^6$Li+$^{208}$Pb at the 
incident energies indicated by the labels. The  dotted, dashed and 
solid lines correspond to the NEB (IAV model), EBU (CDCC) and their sum (TBU), 
respectively. Experimental data 
are from Refs.~\cite{Sig01,Kelly00}. See text for details.}
\end{center}
\end{figure}
%-------------------------------------------------------------
First, the results for the reaction $^{208}$Pb($^6$Li,$\alpha X$), at several 
energies between 29 and 39 MeV are presented, comparing with the data from Refs.~\cite{Sig01,Kelly00}. 
The nominal Coulomb barrier for this system is around 29.5 MeV \cite{Sig01}. The 
CDCC calculations use the same structure model and bin discretization as 
in our previous calculations for  $^6$Li+$^{209}$Bi \cite{Jin15}. The $d-^{208}$Pb and $\alpha-^{208}$Pb optical 
potentials are taken from Refs.~\cite{Han06} and \cite{Barnett74}, 
respectively. To improve the reproduction of the elastic data, the surface term of the imaginary 
part of the $d+^{208}$Pb potential was removed. For the NEB calculations, 
the optical potential of $^6$Li+$^{208}$Pb is taken from Ref.~\cite{Cook82}.

Figure \ref{fig:li6pb_dsdw} shows the comparison of the calculated and 
experimental angular distributions of $\alpha$ particles produced in 
this reaction at the measured incident energies. 
The squares and circles are the experimental data from 
Refs.~\cite{Sig01} and \cite{Kelly00}, respectively. It is evident
that there is an appreciable difference between the two sets of data.
The dashed and dotted lines are the EBU (CDCC) and NEB (IAV model) results. As in the 
$^6$Li+$^{209}$Bi case \cite{Jin15}, the NEB is found to account for most of the inclusive breakup 
cross section. The sum EBU+NEB (TBU) reproduces reasonably well the magnitude and shape 
of the data of Ref.~\cite{Sig01}, except for some overestimation for the lowest energies. Thus, our calculations clearly favour the data 
presented in Ref.~\cite{Sig01} over those presented in \cite{Kelly00}.

From the results shown here and in Ref.~\cite{Jin15}, it 
can be concluded that the nonelastic breakup 
process is the dominant $\alpha-$emitting channel 
in the $^6$Li induced reactions on heavy targets. To investigate 
whether this conclusion is a general feature of $^6$Li induced reactions 
or it holds only for heavy targets we extend our analysis to lighter targets.

%----------------------------------------------------------------
\subsection{$^{159}$Tb ($^6$Li, $\alpha X$)\label{sec:6li159tb}}
%----------------------------------------------------------------
This reaction has been measured by 
Pradhan \textit{et al.} \cite{Pra13} at several energies between 23 MeV and 35 MeV. 

In Ref.~\cite{Pra13}, the following processes were invoked to explain the observed $\alpha$ yields: 
(i) breakup of $^6$Li into $\alpha$ and $d$ fragments where both 
fragments escape without being captured by the target, referred to in some works as  non-capture 
breakup; (ii) $\alpha$ particles resulting from $d$ capture 
by the target ({\it deuteron incomplete fusion}), following the breakup of $^6$Li into $\alpha$ 
and $d$ or a deuteron transfer to the target; (iii) 
single-proton stripping from $^6$Li to produce the unbound $^5$He nucleus that decays 
into an $\alpha$ particle and a neutron; (iv) single-neutron stripping from 
$^6$Li to produce $^5$Li, which will subsequently decay 
into an $\alpha$+p; and (v) single-neutron pickup 
from $^6$Li to produce $^7$Li, which breaks into an $\alpha$ particle 
and a triton if $^7$Li is excited above its breakup threshold of 2.468 MeV. In Ref.~\cite{Pra13} these processes were treated separately, using several reaction formalisms and their sum reasonably reproduced  the total  $\alpha-$particle cross 
sections, but not their angular distributions. 

Within the inclusive breakup model adopted here, 
the processes discussed by Pradhan \textit{et al.}~\cite{Pra13} 
can be re-defined as follows:  process (i) can be divided into 
two parts. First, the non-capture breakup with the target 
remaining in its ground state, i.e., EBU. Second, the non-capture breakup 
accompanied by target excitation, which we call {\it inelastic breakup} and 
is part of our {\it non-elastic breakup} cross section; 
processes (ii)-(iv) may be also embedded in the NEB part, in which the deuteron 
is absorbed by the target or it breaks up into $p+n$ following 
the breakup of $^6$Li into $\alpha$ and $d$; 
it can also happen that after the breakup of $^6$Li, the deuteron picks 
a neutron to become a tritium, contributing to the process (v). 
Processes (ii)-(v) as well as the inelastic breakup can be considered as
nonelastic breakup and should be therefore accounted by the IAV formalism. 

%-------------------------------------------------------------
\begin{figure}[]
\begin{center}
 {\centering \resizebox*{0.85\columnwidth}{!}{\includegraphics{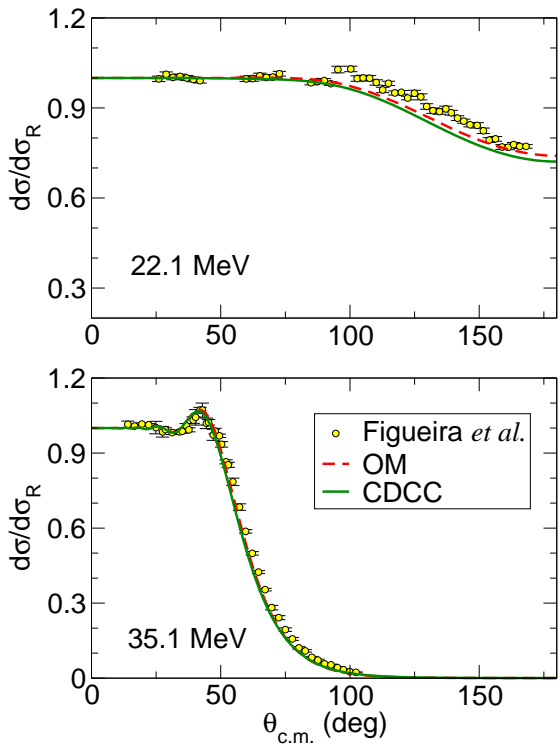}} \par}
\caption{\label{fig:6li144sm}(Color online) Elastic scattering of $^6$Li+$^{144}$Sm at 22.1 MeV (top) and 35.1~MeV (bottom). The solid and dashed  lines are, 
respectively, the CDCC calculation and the optical model calculation with the 
optical potential from \cite{Cook82}. Experimental data are from Ref.~\cite{Figueira10} .}
\end{center}
\end{figure}
%-------------------------------------------------------------

Elastic data for this reaction are not available. Thus, the CDCC 
calculation  is tested against the data for the nearby system $^6$Li+$^{144}$Sm \cite{Figueira10}. 
The $\alpha +^{144}$Sm and $d +^{144}$Sm optical potentials were taken from Refs.~\cite{Hui62} and \cite{Han06}, respectively.  The optical model calculation using the potential of 
Cook \cite{Cook82} (dashed lines) is also shown. It can be seen that the 
CDCC result is similar to the optical model calculation, particularly at 
$E=35.1$ MeV. At this energy, the calculations reproduce very well the elastic 
data. For the lower energy ($E=22.1$~MeV), both calculations underestimate the data at 
backward angles. Note that, in contrast to the $^6$Li+$^{208}$Pb case, no apparent modification of the deuteron potential was required in this case. 

\begin{comment}
First the elastic scattering with the CDCC framework is studied. Since there  are no elastic scattering data available for the $^6$Li+$^{159}$Tb system, the CDCC calculations are adjusted to reproduce the elastic scattering data of a similar system, i.e., $^6$Li+$^{144}$Sm\cite{Figueira10}. 
The same interaction of $\alpha - d$ as discussed in the previous cases is 
used. The optical potentials of $\alpha-^{144}$Sm and $d-^{144}$Sm were 
evaluated at $2/3$ and $1/3$ of the incident energy of $^6$Li, respectively. 
The global optical model potential parameters from Refs.\cite{Hui62,Han06} were 
used to describe the interactions at the corresponding energies. The 
CDCC calculation is shown in Fig.~\ref{fig:6li144sm} by solid lines. 
For comparison, the optical model calculation using the potential of 
Cook\cite{Cook82} (dashed lines) is also shown. It can be seen that the 
CDCC result is similar to the optical model calculation, particularly at 
$E=35.1$ MeV. At this energy, the calculations reproduce very well the elastic 
data. For the lower energy, the calculations underestimate the data at 
backward angles. Note that, in contrast to the $^6$Li reactions
on heavy targets, i.e., $^6$Li+$^{209}$Bi (Sec.~\ref{sec:6li209bi}) and 
$^6$Li+$^{208}$Pb (Sec.~\ref{sec:6li208pb}), the surface term of the imaginary part of  
the $d-$target potential was kept in this case.
\end{comment}

%-------------------------------------------------------------
\begin{figure}[]
\begin{center}
 {\centering \resizebox*{0.9\columnwidth}{!}{\includegraphics{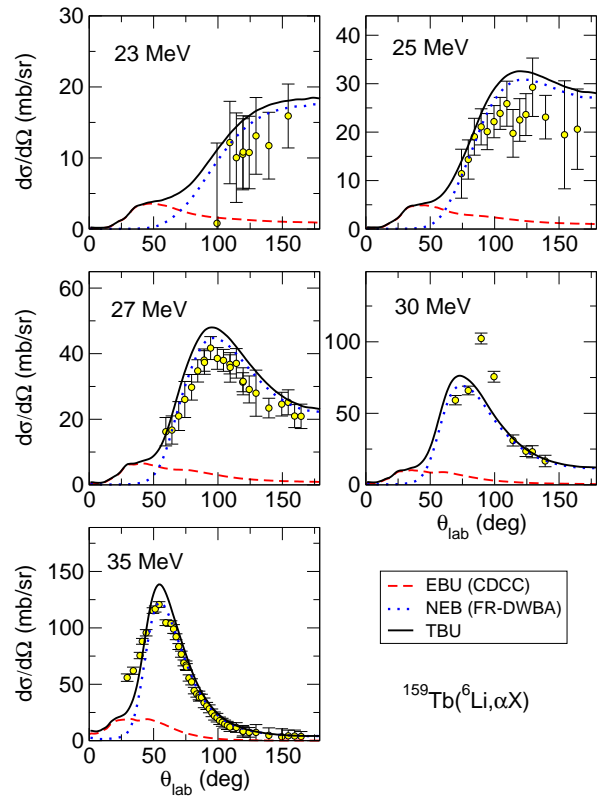}} \par}
\caption{\label{fig:6li159tb}(Color online) Angular distribution of $\alpha$ particle 
production of the reaction $^6$Li+$^{159}$Tb at the incident energies indicated by 
the labels. The dashed, dotted and solid lines are EBU calculated with CDCC, 
NEB calculated with finite-range DWBA and their sum (TBU), respectively. The experimental
data are taken from Ref.~\cite{Pra13}}
\end{center}
\end{figure}
%-------------------------------------------------------------

Now the inclusive breakup cross sections $^{159}$Tb($^6$Li,$\alpha X$) 
are discussed. The EBU contribution was obtained from the CDCC calculations 
discussed in the previous paragraph. For the NEB calculation, 
the same optical potentials $\alpha/d+^{159}$Tb were 
used. The Cook potential \cite{Cook82} was used to calculate 
the distorted wave of the incoming channel.  

In Fig.~\ref{fig:6li159tb} the calculated and experimental angular distributions
of $\alpha$ particles are compared for several incident energies of $^6$Li. 
The dashed and dotted lines are the EBU (CDCC) and NEB (IAV model) results. 
The summed EBU + NEB cross sections  (solid lines) reproduce fairly well the shape and magnitude of the data, 
except for a slight overestimation at some energies.
Similarly to the heavy-target systems, i.e., $^6$Li+$^{209}$Bi \cite{Jin15} and 
$^6$Li+$^{208}$Pb (Sec.~\ref{sec:6li208pb}), the NEB is found to account for most of the 
inclusive breakup cross section.

%-------------------------------------------------------------
\subsection{$^{118}$Sn ($^6$Li, $\alpha X$)\label{sec:6li118sn}}
%-------------------------------------------------------------
% 6Li+118Sn breakup ds/dw
\begin{figure}[tb]
\begin{center}
 {\centering \resizebox*{0.95\columnwidth}{!}{\includegraphics{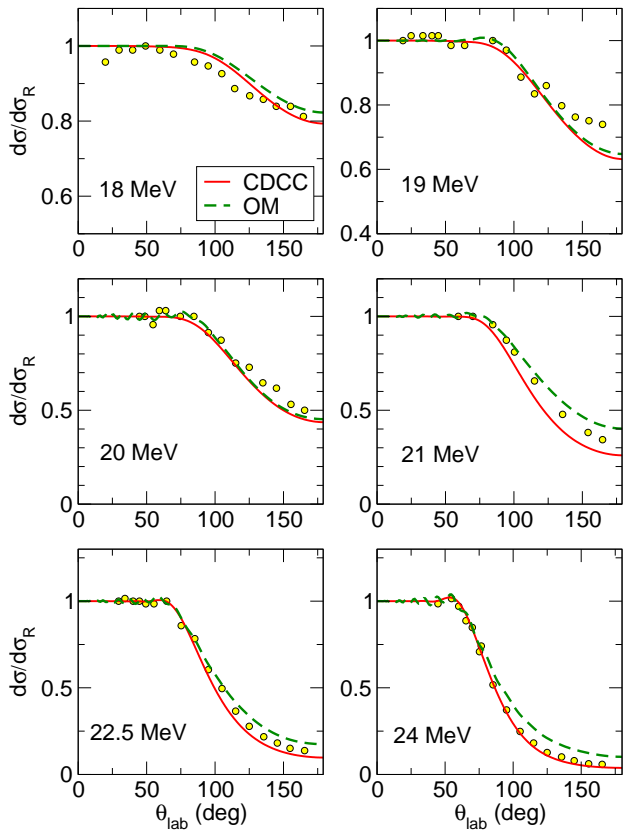}} \par}
\caption{\label{fig:6li118snel}(Color online) Elastic scattering of $^6$Li+$^{118}$Sn at 
different incident energies. The solid and dashed lines are, 
respectively, the CDCC calculation and the optical model calculation with 
the optical potential from \cite{Pfeiffer73}. 
Experimental data are from Ref.~\cite{Pfeiffer73}.}
\end{center}
\end{figure}
%-------------------------------------------------------------
Inclusive breakup data for the $^{118}$Sn($^6$Li, $\alpha X$) reaction are available in Ref.~\cite{Pfeiffer73} at energies 
between 18 and 24 MeV. The optical 
model parameterizations of Refs.~\cite{Hui62} and \cite{Han06} are used for the $\alpha-^{118}$Sn
and $d-^{118}$Sn systems. For the NEB 
calculations, the optical potential of $^6$Li+$^{118}$Sn is taken from 
Ref.~\cite{Pfeiffer73}. 

In Fig.~\ref{fig:6li118snel} we compare the elastic data with the CDCC (solid lines) and optical model  (dashed lines) calculations. Overall, both types of calculations reproduce 
well the data, with some discrepancy observed at 18 and 21 MeV.   

%First the validity of the two-cluster model of $^6$Li was studied for the elastic  scattering of $^6$Li + $^{118}$Sn at several incident energies. It is 
%found that the CDCC calculations with this model are in good agreemen
%with the experimental data. This is shown in Fig.~\ref{fig:6li118snel} by
%solid lines. For comparison, the optical model calculation using the 
%potential mentioned in Ref.~\cite{Pfeiffer73} (dashed lines) is also shown. 

%-------------------------------------------------------------
% 6Li+118Sn breakup ds/dw
\begin{figure}[tb]
\begin{center}
 {\centering \resizebox*{0.95\columnwidth}{!}{\includegraphics{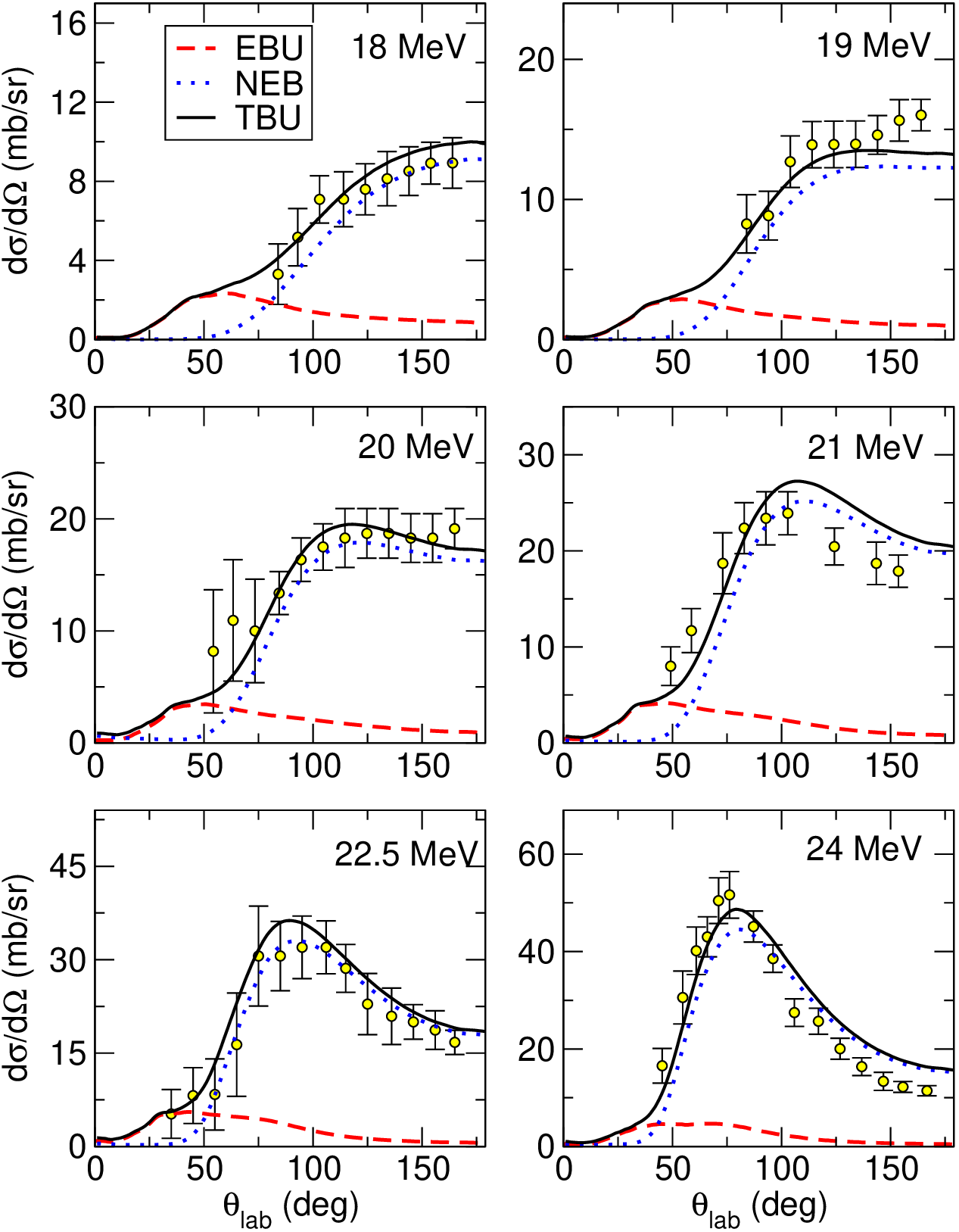}} \par}
\caption{\label{fig:6li118sn}(Color online) Angular distribution of 
$\alpha$ particles produced in the reaction $^6$Li+$^{118}$Sn at the 
incident energies indicated by the labels. The  dotted, dashed and 
solid lines correspond to the NEB (IAV model), EBU (CDCC) and their sum (TBU), 
respectively. Experimental data are from Ref.~\cite{Pfeiffer73}.}
\end{center}
\end{figure}
%-------------------------------------------------------------
Figure \ref{fig:6li118sn} shows the comparison of the calculated and 
experimental angular distributions of $\alpha$ particles produced in this 
reaction, for several incident energies. 
%The dashed and solid lines denote the EBU (CDCC) and NEB results. 
 Again, the  
NEB part (dotted lines) accounts for most of the inclusive breakup cross section and the 
EBU (dashed lines) becomes the dominant breakup mode for angles smaller than $\sim$50 degrees. The 
summed EBU  + NEB result (solid line) reproduces remarkably well the shape and 
magnitude of the data. 

%We have seen that the IAV model works rather well for $^6$Li reactions with  heavy-mass and medium-mass targets. In the following subsections, we examine the  validity of the model for lighter targets. 

%-------------------------------------------------------------
\subsection{$^{59}$Co ($^6$Li, $\alpha X$)\label{sec:6li59co}}
%-------------------------------------------------------------
% 6Li+59Co elastic ds/ds_R
\begin{figure}[tb]
\begin{center}
 {\centering \resizebox*{0.85\columnwidth}{!}{\includegraphics{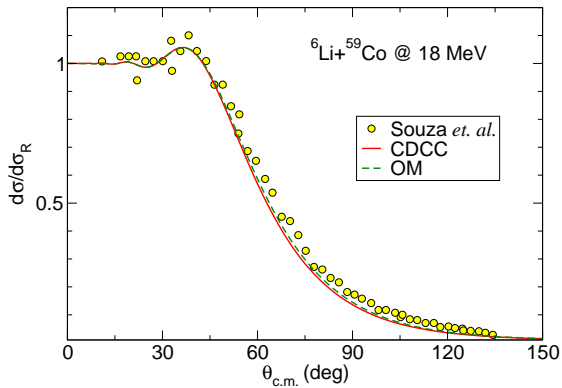}} \par}
\caption{\label{fig:6li59coel}(Color online) Elastic scattering of $^6$Li+$^{59}$Co at an
incident energy of $18$ MeV. The solid and dashed lines are, 
respectively, the CDCC calculation and the optical model calculation with 
the optical potential from~\cite{Cook82}. 
Experimental data are from Ref.~\cite{Souza07}.}
\end{center}
\end{figure}
%-------------------------------------------------------------

%-------------------------------------------------------------
% 6Li+59Co breakup ds/dw
\begin{figure}[tb]
\begin{center}
 {\centering \resizebox*{0.85\columnwidth}{!}{\includegraphics{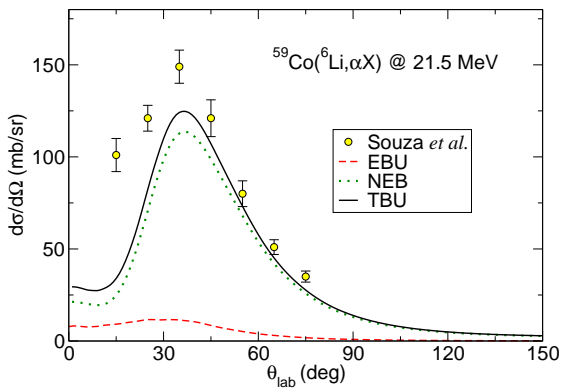}} \par}
\caption{\label{fig:6li59codsdw}(Color online) Angular distribution of $\alpha$ particles 
produced in the reaction $^6$Li+$^{59}$Co at an incident energy of $21.5$ MeV. 
The dashed, dotted, and solid lines are, respectively, the EBU (CDCC), 
NEB(IAV model) and their sum. Experimental data are taken from Ref.~\cite{Souza09}.}
\end{center}
\end{figure}
%-------------------------------------------------------------
%-------------------------------------------------------------
% 6Li+59Co breakup ds/de
\begin{figure}[tb]
\begin{center}
 {\centering \resizebox*{0.85\columnwidth}{!}{\includegraphics{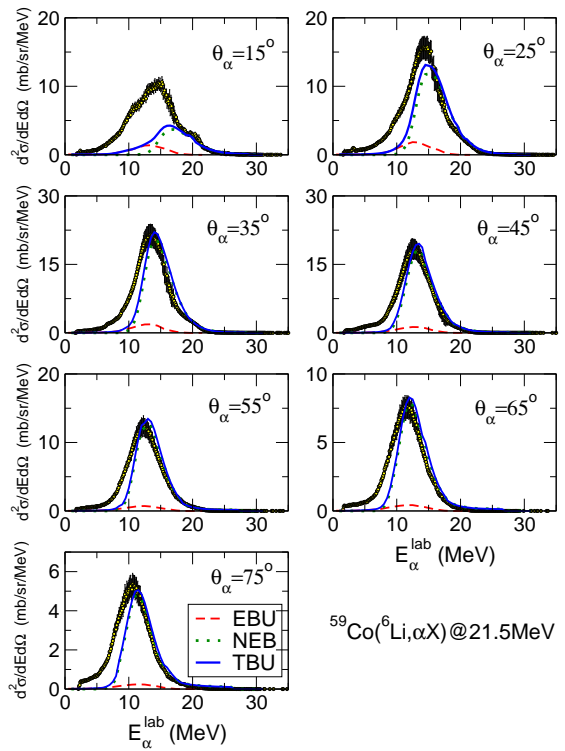}} \par}
\caption{\label{fig:6li59codsde}(Color online) Experimental and calculated inclusive
$\alpha$ energy spectra for $E_{\mathrm{lab}}=21.5$ MeV, at selected scattering angles. 
%$\theta_\mathrm{lab}=15$, $25$, $35$, $45$, $55$, $65$ and $75$ degrees. 
The dashed, dotted, and solid lines are respectively the EBU (CDCC), 
NEB (IAV model) and their sum. Experimental data are taken from Ref.~\cite{Souza09}.}
\end{center}
\end{figure}
%-------------------------------------------------------------
%As discussed before, for the heavy targets $^{209}$Bi  and $^{208}$Pb and medium-heavy targets 
% $^{159}$Tb and $^{118}$Sn the nonelastic breakup process dominates the 
%inclusive $\alpha$ production in the $^6$Li induced reactions. 
Experimental data for the  $\alpha$-production channel for the reaction $^6$Li+$^{59}$Co  have been reported
 by Souza \textit{et al.}~\cite{Souza09} at  $E_{\mathrm{lab}}=21.5$ MeV, which is above the Coulomb barrier ($V_B=12$ MeV). 

Elastic data are available at the somewhat smaller energy $E_{\mathrm{lab}}=18$ MeV \cite{Souza07} so we first compare these data with the optical 
model and CDCC calculations. For the former, we employed the global optical potential of Cook \cite{Cook82}. For the CDCC calculations, 
the optical potentials for 
$\alpha+^{59}$Co and $d+^{59}$Co were taken from Refs.~\cite{Hui62} and 
\cite{Han06}, respectively. The results are shown in Fig.~\ref{fig:6li59coel}.  
It can be seen that both the CDCC and optical model calculations reproduce fairly 
well the  data. 
We notice that no renormalization of the deuteron potential was required
in this case. 

The experimental and calculated angular distributions of inclusive $\alpha$ particles are shown in Fig.~\ref{fig:6li59codsdw}. The NEB is seen to dominate the inclusive 
$\alpha$ production. 
It should be noticed that, in this case, the NEB part includes also the 
transfer populating bound states of the target, which was obtained  
using the formalism discussed in Sec.~\ref{sec:ex0}. A more detailed discussion 
of this contribution is left for Sec.~\ref{sec:nebvstr}.
The total cross section, TBU= EBU + NEB, reproduces well
the shape of the experimental data, although the magnitude is underestimated by 
$\sim$30\% at the maximum. This might indicate the presence of  other 
relevant mechanisms leading to the production of $\alpha$ particles in this reaction, such as the formation of a compound nucleus followed by $\alpha$ evaporation.  In fact, statistical model calculations performed in Ref.~\cite{Souza09} predicted a significant amount of $\alpha$ particles coming from this channel. The evaluation of this contribution is beyond the scope of the present work. 

The  energy spectra for selected $\alpha$ scattering angles are  also available for this reaction. These are compared with our calculations in  Fig.~\ref{fig:6li59codsde}, with each panel corresponding to a given $\alpha$ scattering angle, as indicated by the labels. 
Except at $\theta_\mathrm{lab}=15^\circ$, the sum of EBU and NEB reproduces
the peak of the $\alpha$ energy distribution.
However, the low-energy tail is clearly underestimated. At these energies, the
main contribution of the inclusive $\alpha$ production may arise from 
compound nucleus followed by evaporation and pre-equilibrium, which are not
considered in the present calculations.  
 We note that high energy $\alpha$ particles stem from a deuteron 
transfer mechanism to the target and  are well reproduced by our calculations. 
%However, further calculations are still planned to investigate these possibilities.

\subsection{$^{58}$Ni ($^6$Li, $\alpha X$)\label{sec:6li58ni}}
%-------------------------------------------------------------
% 6Li+58Ni elastic ds/ds_R
\begin{figure}[tb]
\begin{center}
 {\centering \resizebox*{0.95\columnwidth}{!}{\includegraphics{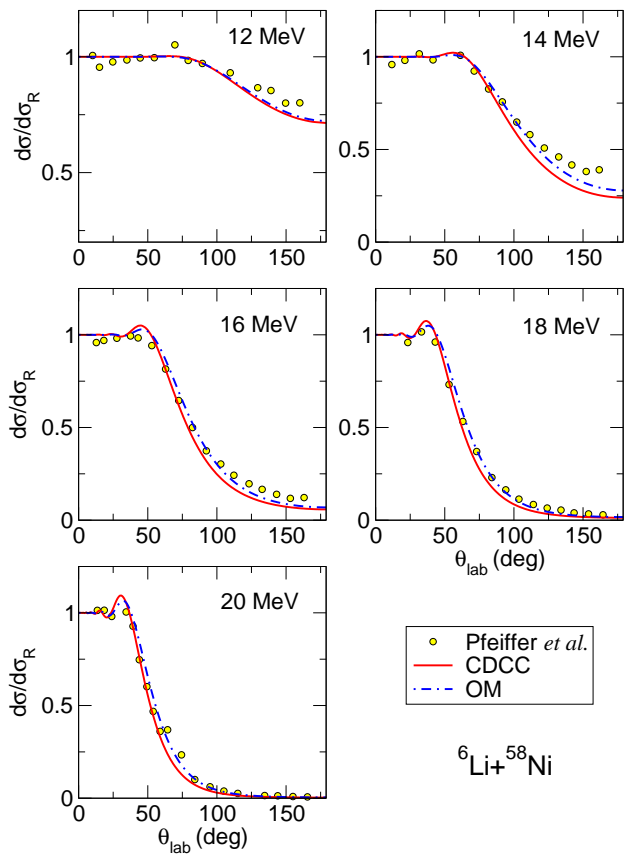}} \par}
\caption{\label{fig:6li58niel}(Color online) Elastic scattering of $^6$Li+$^{58}$Ni at 
several energies indicated by the labels. The solid and dashed lines are, 
respectively, the CDCC calculation and the optical model calculation with 
the optical potential from~\cite{Pfeiffer73}. 
Experimental data are from Ref.~\cite{Pfeiffer73}.}
\end{center}
\end{figure}
%-------------------------------------------------------------
%-------------------------------------------------------------
% 6Li+58Ni elastic ds/ds_R
\begin{figure}[tb]
\begin{center}
 {\centering \resizebox*{0.95\columnwidth}{!}{\includegraphics{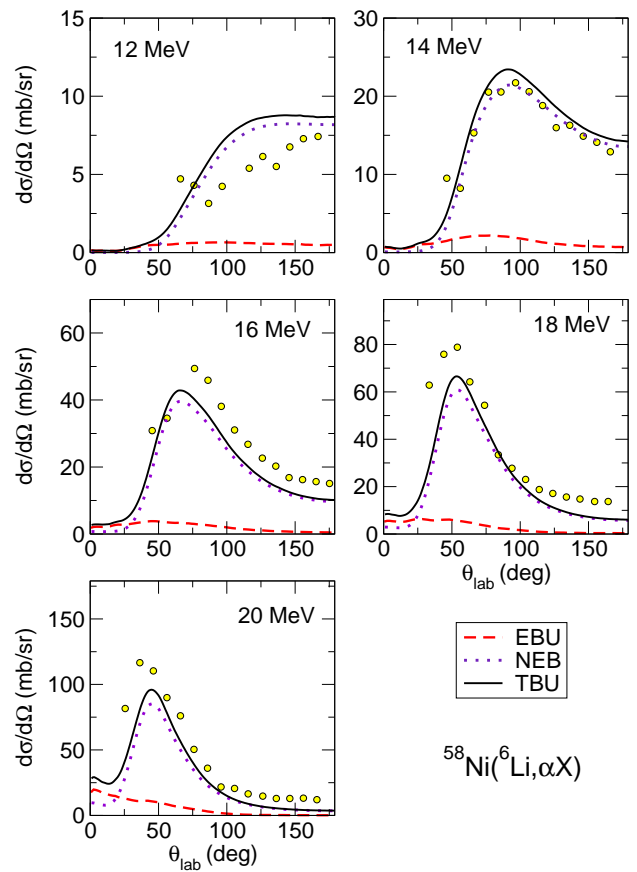}} \par}
\caption{\label{fig:6li58ni}(Color online) Angular distribution of $\alpha$ particles
produced in the reaction $^6$Li + $^{58}$Ni at the incident energies 
indicated by the labels. The dashed, dotted and solid lines are, 
respectively, the EBU, NEB and their sum (TBU). 
Experimental data are from Ref.~\cite{Pfeiffer73}.}
\end{center}
\end{figure}
%-------------------------------------------------------------

The $\alpha$ production of the $^6$Li + $^{58}$Ni reaction at several incident
energies between 12 MeV and 20 MeV was measured by Pfeiffer \textit{et al.}
\cite{Pfeiffer73}. Elastic scattering data,  which were also measured, are compared with CDCC 
and OM calculations in Fig.~\ref{fig:6li58niel} (note that the angles and cross sections are referred to the laboratory frame, as in the original reference). For the former, we use the same optical potentials as in the nearby $^6$Li+$^{59}$Co case. For the OM calculations we use the global OM potential by Cook \cite{Cook82}.
%To test the two-body structure of $^6$Li and the fragment-target interactions, the elastic scattering data were  compared with CDCC calculations. The same optical potentials as  in the $^6$Li+$^{59}$Co case were used. 
Both calculations reproduce rather well the data, although the CDCC calculations slightly underestimates the data at large angles. 

We present now the inclusive alpha cross sections.
For the NEB calculation, the $^6$Li optical potential from 
Ref.~\cite{Cook82} was used.  Figure  \ref{fig:6li58ni} shows the comparison
of the calculated and experimental angular distributions of $\alpha$ 
particles produced in this reaction, for several incident 
energies.
%The dashed and dotted lines are the EBU and NEB contributions obtained with the CDCC and IAV formalisms, respectively. 
Notice that the NEB (dotted lines) includes also the contribution coming from the  transfer to target bound states. 
Again, the NEB part dominates the inclusive $\alpha$ production. In general, the summed EBU + NEB cross section (solid lines) 
reproduces  well the shape and magnitude of the data. At $16$,  $18$ and $20$ MeV some underestimation is observed, which might be  associated with other $\alpha$-production channels, as pointed out in the $^6$Li+$^{59}$Co case.

From the results presented in the previous sections, we may conclude that 
the strong $\alpha$-production channel observed in $^6$Li experiments 
originates mostly from non-elastic breakup mechanisms. In all cases analyzed, the EBU mode turns out to account for a relatively small fraction 
of the total inclusive alpha cross section and its contribution is only 
important for the alpha particles emitted at small angles.  For the lighter targets, we found also 
a indirect evidence of other alpha production mechanisms, such as 
fusion.

%--------------------------------------------------------------------------

%\section{NEB Vs. transfer}\label{sec:nebvstr}
\section{Transfer content of the NEB cross section}\label{sec:nebvstr}

% NEB with different incoming energies 
\begin{figure}[tb]
\begin{center}
 {\centering \resizebox*{0.85\columnwidth}{!}{\includegraphics{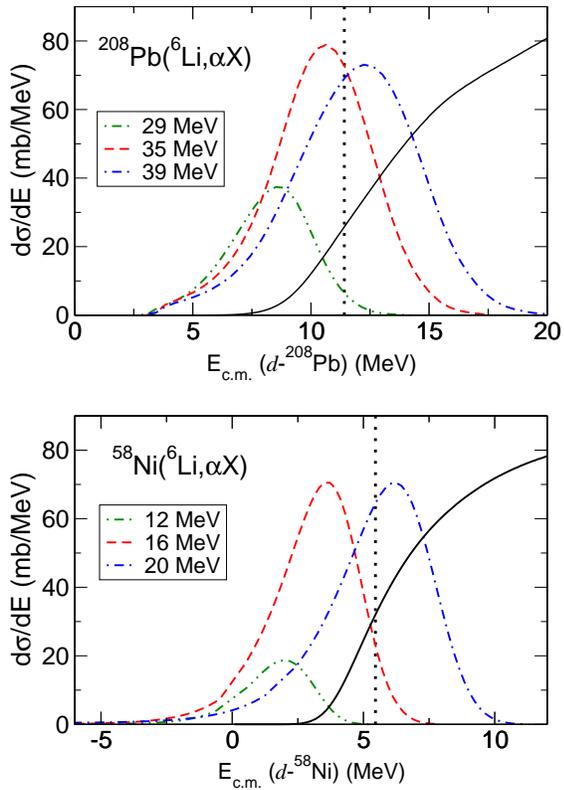}} \par}
\caption{\label{fig:NEBex}(Color online) Top: NEB cross section as a function of the 
$d$--$^{208}$Pb relative energy in the c.m.\ frame for the reaction
 $^6$Li+$^{208}$Pb. The vertical dotted line indicates
the energy of the Coulomb barrier for the  $d$+$^{208}$Pb reaction. The solid line is the reaction cross section for $d$+$^{208}$Pb, arbitrarily normalized. Bottom: same as in top panel but for the  $^6$Li+$^{58}$Ni system. } 
\end{center}
\end{figure}
% NEB vs transfer 
\begin{figure}[tb]
\begin{center}
 {\centering \resizebox*{0.95\columnwidth}{!}{\includegraphics{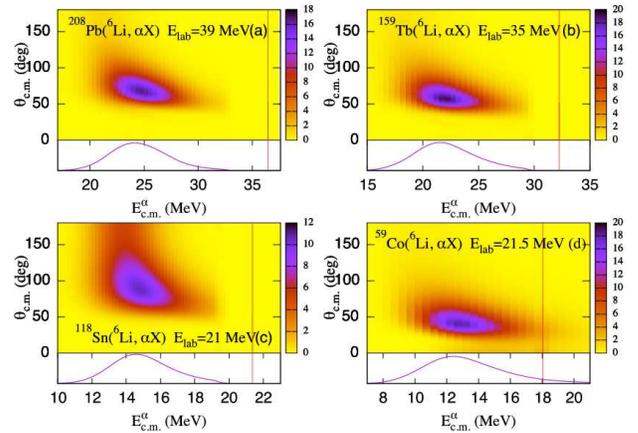}} \par}
\caption{\label{fig:neb_transfer}(Color online) Contour plots for the  double differential 
cross section (upper panels) and the angle-integrated enery differential cross section  as a function of the
outgoing $\alpha$ energy in the c.m.\ frame (lower panels) for the reactions:   
(a) $^6$Li+$^{208}$Pb, (b) $^6$Li+$^{159}$Tb, (c) $^6$Li+$^{118}$Sn and 
(d) $^6$Li+$^{59}$Co. The vertical lines indicate the breakup threshold for the $d$+target
system ($E_x=0$). }
\end{center}
\end{figure}

\begin{comment}
The appreciable difference between the NEB and transfer to the 
target bound state is the the relative energy.
The NEB is the one with a positive energy between the deuteron and target, 
whereas the transfer has a negative relative energy. In IAV model, the NEB combines
all the other possible reaction channels expect the elastic scattering 
between $d$ and target, which is the loss of flux leaving the elastic 
channel and included in the imaginary pat of optical potential. 
On the other hand, imaginary part used in extended-IAV model for the transer
to the target bound state is associated with the spreading width of the 
single-particle states, which accounts for the fragmentation of these 
states into more complicated configurations. 
\end{comment}

The relative importance of the transfer to bound states within the NEB cross section will depend on several parameters, 
such as the projectile incident energy a
nd the charge of the target nucleus. 
For heavy targets, the transfer channel is suppressed due to the 
strong Coulomb interaction between the deuteron and the target, whereas for light targets this channel is expected to play a more important role.

This is illustrated in Fig.~\ref{fig:NEBex} for two such cases; the upper panel displays the 
calculated $^{208}$Pb($^6$Li, $\alpha$X) NEB cross sections as a function of 
$d$-$^{208}$Pb relative  energy  
at three different incident energies, 
29 MeV, 35 MeV and 39 MeV. The vertical dotted line indicates
the nominal Coulomb barrier for the $d$-$^{208}$Pb system. The black solid curve is the reaction cross section for the $d$-$^{208}$Pb system, arbitrarily normalized to fit within the same scale. The bottom panel shows similar curves  for the $^{6}$Li+$^{58}$Ni reaction at 12, 16 and 20 MeV.  In both cases, it can be seen that the 
NEB is a \textit{Trojan Horse} type process \cite{Bau86},  which means that the $^6$Li projectile 
brings the deuteron inside the Coulomb barrier and let it interact with the target nucleus, giving a sizable cross sections for deuteron energies for which the reaction cross section has already become negligibly small.  
For the $^{208}$Pb target, due to the strong Coulomb repulsion, the NEB cross section becomes 
negligible at negative $d$-$^{208}$Pb relative energies and this behavior is 
independent of the incoming $^6$Li energy. By contrary, for the $^{58}$Ni target, there is a low energy tail extending to negative deuteron 
energies (transfer). 
%For both targets one can see that the deuteron-target reaction cross section vanishes much faster than the ($^6$Li,$\alpha$) cross section

We expect also some correlation between the $\alpha$-particle angular and energy distribution. This is shown in Fig.~\ref{fig:neb_transfer} in the form of contour plots of double differential cross sections and angle-integrated cross section as a function of the outgoing $\alpha$ energy in the c.m.\ frame for the reactions  
(a) $^6$Li+$^{208}$Pb, (b) $^6$Li+$^{159}$Tb, (c) $^6$Li+$^{118}$Sn and 
(d) $^6$Li+$^{59}$Co. It can seen that the most energetic $\alpha$  particles are preferably emitted at forward angles, whereas those with lower energies contribute to both forward and backward angles. Moreover, 
when the charge of the target is small ($^{59}$Co), the transfer 
channel becomes more relevant.

%-------------------------------------------------------------------------
\section{Systematics of inclusive $\alpha$ production \label{sec:6lisys}}
%-------------------------------------------------------------------------

%-------------------------------------------------------------
% 6Li systematics of inclusive $\alpha$ 
\begin{figure}[tb]
\begin{center}
 {\centering \resizebox*{0.95\columnwidth}{!}{\includegraphics{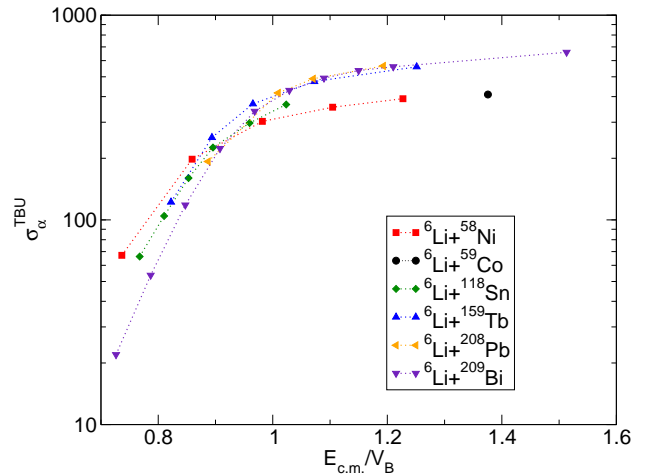}} \par}
\caption{\label{fig:6lisys}(Color online)  Inclusive breakup $\alpha$ cross sections 
involving $^6$Li projectile with several targets as a function 
of $E_\mathrm{c.m.}/V_\mathrm{B}$. 
%(b) Reduced inclusive breakup $\alpha$ cross sections as a function of $E_\mathrm{c.m.}/V_\mathrm{b}$ for the 
%same systems as (a). See text for the details.}
}
\end{center}
\end{figure}
%-------------------------------------------------------------

%-------------------------------------------------------------
% 6Li systematics of inclusive $\alpha$ 
\begin{figure}[t]
\begin{center}
 {\centering \resizebox*{0.95\columnwidth}{!}{\includegraphics{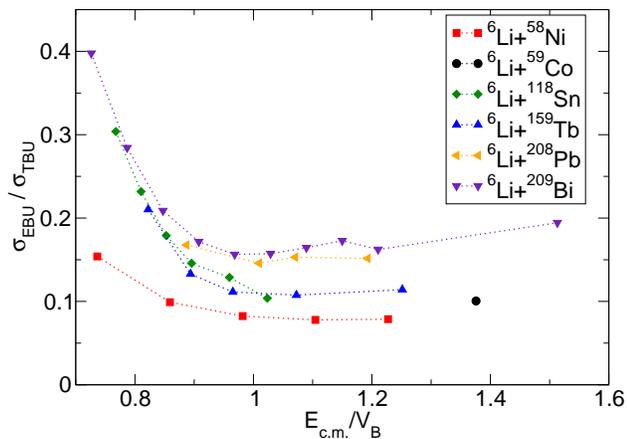}} \par}
\caption{\label{fig:eburatio}(Color online) Ratios of calculated EBU over 
TBU (= EBU + NEB) for different systems. See text for details.}
\end{center}
\end{figure}
%-------------------------------------------------------------

Systematic studies of $\alpha$ production yields in $^6$Li reactions 
show an interesting universal behaviour when plotted as a function of 
the incident energy scaled by the Coulomb
barrier energy as reported for instance by Pakou \textit{et al.}~\cite{Pak03}. In this section, we 
will investigate whether our calculations exhibit also this universal behaviour. 
For this study, we have considered the target systems $^{59}$Co, $^{118}$Sn, $^{159}$Tb,
$^{208}$Pb, which have been analyzed in the preceding sections, and  $^{209}$Bi, analyzed in Ref.~\cite{Jin15}. 
The results are shown in Fig.~\ref{fig:6lisys},
where we plot the calculated $\sigma_\alpha^\mathrm{TBU}$ cross sections
as a function of the reduced energy 
($E_\mathrm{c.m.}/V_\mathrm{B}$), with $V_\mathrm{B}$ the energy of the
Coulomb barrier, estimated as $V_\mathrm{B}=Z_pZ_te^2/(r_B(A_p^{1/3}+A_t^{1/3}))$,
where $Z_p$ ($Z_t$) and $A_p$ ($A_t$) are the atomic number and atomic mass 
of the projectile (target), respectively, and $r_B=1.44$ fm. 
%The squares, circles, diamonds, up triangles, left triangles and down triangles correspond, respectively, to the reactions of 
%$^6$Li + $^{58}$Ni, $^6$Li + $^{59}$Co, $^6$Li + $^{118}$Sn, $^6$Li + $^{159}$Tb, $^6$Li + $^{208}$Pb and $^6$Li + $^{209}$Bi.  
As expected, the breakup cross section drops quickly as the incident energy decreases below the barrier. This effect is enhanced for the 
 $^{209}$Bi nucleus, possibly due to the larger Coulomb repulsion. 
Above the barrier,  the medium-heavy and 
heavy targets the inclusive breakup cross sections 
show a similar trend, but not for the medium mass targets $^{58}$Ni and $^{59}$Co at larger energies. 
We recall however that, for these lighter systems, there might be 
additional contributions from other channels, such as compound nucleus 
following evaporation, which are not accounted for by the IAV formalism.

%Fig.~\ref{fig:6lisys} (b) shows the reduced inclusive breakup $\alpha$  cross sections, $\sigma_\alpha^\mathrm{TBU}/(\pi R_B^2)$ with $R_B=r_B(A_p^{1/3}+A_t^{1/3})$, as a function of the reduced energy. The curves of reduced inclusive breakup cross sections are clearly different for medium, medium-heavy and heavy mass targets. For a given reduced energy, the reduced inclusive breakup cross section decreases with the product $Z_pZ_t$, in agreement with the calculations of Ref.\cite{Canto15}.

% 6Li systematics of inclusive $\alpha$ 
\begin{figure}[t]
\begin{center}
 {\centering \resizebox*{0.95\columnwidth}{!}{\includegraphics{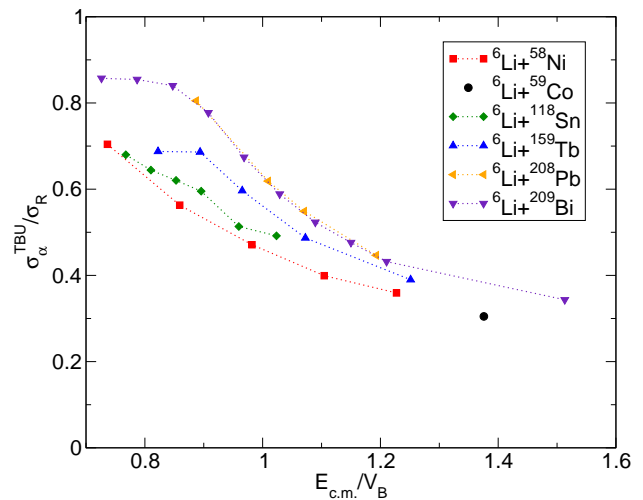}} \par}
\caption{\label{fig:ratioreac}(Color online) Ratios of calculated TBU (= EBU + NEB) $\alpha$ cross sections over the reaction cross section for the systems and energies analyzed in this work. See text for details.}
\end{center}
\end{figure}
%-------------------------------------------------------------

We have also studied the relative importance of EBU versus NEB as a function
of the incident energy. For that, we display in 
Fig.~\ref{fig:eburatio} the ratio of EBU over TBU (= EBU + NEB) for the analyzed systems.
It is seen that, for incident energies below 
the Coulomb barrier, the  elastic breakup cross section becomes comparatively more important as  
the energy decreases. This can be attributed to the fact that, below the barrier, the breakup takes place at large projectile-target separations, and the deuteron absorption (responsible for the NEB part) will be less important \cite{Jin16}. By contrast, for energies above the Coulomb barrier, 
the ratio shows an almost constant behavior. 
It can also be seen that, while for the heavy mass targets elastic 
breakup plays an important role in the inclusive $\alpha$ production, 
especially below the Coulomb barrier, for the medium mass targets elastic 
breakup is less important and the nonelastic breakup is dominant.  

Another relevant question regards the fraction of the reaction cross section that is exhausted by the $\alpha$ cross section. To address this question, 
we plot in Fig.~\ref{fig:ratioreac} the ratio of the calculated TBU and reaction cross sections as a function of the reduced energy $E_\mathrm{c.m.}/V_B$, for the  systems studied in this work. Several interesting features emerge from this plot: (i) first, for all systems analyzed the ratio decreases smoothly as the incident energy increases; (ii) second, the heavier the target nucleus, the larger the ratio. For example, for the $^{208}$Pb and $^{209}$Bi target nuclei the ratio exceeds 80\% at sub-Coulomb energies. Result (i) may be understood as a consequence of the competition with other channels which will open and increase their importance as the incident energy increases, such as other breakup modes not associated with the production of $\alpha$ particles (e.g. $^3$H+$^3$He), target excitation not accompanied by projectile breakup, neutron pickup from the target, etc.

%-------------------------------
\section{Summary and conclusions}\label{sec:sum}
%--------------------------------
To summarize, we have performed a comprehensive analysis of inclusive breakup  cross sections  in $^{6}$Li-induced reactions with the aim of understanding the experimentally observed $\alpha$ yields. For that, we have calculated separately the EBU and NEB contributions using the CDCC method (for the EBU part) and the closed-form model proposed by Ichimura, Austern and Vincent IAV theory (for the NEB part). For the latter model, we used the DWBA approximation,  including finite-range effects and the remnant term of the transition operator. 

Overall, the calculations show a very good agreement with the available data, providing a consistent and neat explanation of the large $\alpha$ yields reported over the years for $^6$Li reactions, without the need of evaluating the individual channels contributing to the inclusive  cross section.   
Furthermore, in all cases analyzed, the total $\alpha$ breakup is largely dominated by the NEB part, with the EBU part representing only  a small fraction of the total inclusive cross section. This explains why the CDCC calculations tend to largely underpredict the measured $\alpha$  yields. The EBU becomes only dominant at very small angles, or at energies well below the Coulomb barrier. For the heavy target systems, the $\alpha$ singles cross section accounts for a large fraction of the reaction cross section (above 80\% at sub-Coulomb energies).  
For the lighter mass targets, we found that part of the $\alpha$ yields corresponds to transfer to bound states of the residual nucleus. To account for this contribution, the IAV model has been conveniently extended, following the formalism developed by previous authors \cite{Uda89,Potel:2015eqa}. 

% We have also use the extended-IAV model to calculate the transfer to target bound state which is visible for medium mass target. 

Finally, we have investigated whether our calculations support the observed universal trend of $\alpha$ yields as a function of the reduced incident energy ($E_\mathrm{c.m.}/V_B$). We find that the computed total breakup cross sections (EBU+NEB) exhibit this trend for the heavy targets, but significant deviations have been found for the light
targets. This could indicate that the latter do not obey the universal 
behavior, but we cannot rule out that the deviations are due to the presence of 
additional $\alpha$ production mechanisms, not included in our calculations. 
This problem deserves further investigation.

%
% Acknowledgements -------------------------------------------------------------- 
% Note that in elsevier documentclass style, there is a command 
% \ack for this purpose!!!!
%
\begin{acknowledgments}
We are grateful to Gregory Potel for his guidance in the extension of the IAV model to negative energies. 
This work has been partially supported by the Spanish
 Ministerio de  Econom\'ia y Competitividad and FEDER funds under project 
 FIS2014-53448-C2-1-P  and by the European Union's Horizon 2020 research and innovation program under grant agreement No.\ 654002. 
\end{acknowledgments}

%--------------------------------------------------------------------------------------------------

%% References
%%
%% Following citation commands can be used in the body text:
%% Usage of \cite is as follows:
%%   \cite{key}         ==>>  [#]
%%   \cite[chap. 2]{key} ==>> [#, chap. 2]
%% 

%% References with bibTeX database:

\bibliography{references}
%\bibliography{inclusive,references}
\end{document}